# STAR FORMATION ASSOCIATED WITH THE SNR IC443

Jin-Long Xu[1,2], Jun-Jie Wang[1,2] and Martin Miller[3]

[1] National Astronomical Observatories, Chinese Academy of Sciences, Beijing 100012, China

[2] NAOC-TU Joint Center for Astrophysics, Lhasa 850000, China

[3] I.Institute of Physics, University of Cologne, Cologne, 50937, Germany

xujl@bao.ac.cn

– 2 –

## ABSTRACT

We have performed the submillimeter and millimeter observations in CO lines toward supernova remnant (SNR) IC443. CO molecular shell is well coincident with the partial shell of the SNR detected in radio continuum observations. The broad emission lines and three 1720-MHz OH masers were detected in CO molecular shell. The present observations have provided further evidence in support of the interaction between the SNR and the adjoining molecular clouds (MCs). The total mass of the MCs is $9.26 \times 10^3$ $M_\odot$. The integrated CO line intensity ratio ($R_{I_{CO(3-2)}/I_{CO(2-1)}}$) for the whole molecular cloud is between 0.79 and 3.40. The average value is 1.58, which is much higher than previous measurements of individual Galactic MCs. Higher line ratios imply that shocks have driven into the MCs. We conclude that high $R_{I_{CO(3-2)}/I_{CO(2-1)}}$ is identified as one good signature of SNR-MCs interacting system. Based on *IRAS* Point Source Catalog and 2MASS near-infrared database, 12 protostellar objects and 1666 young stellar objects (YSOs) candidates (including 154 CTTSs and 419 HAeBe stars) are selected. In the interacting regions, the significant enhancement of the number of protostellar objects and YSOs indicates the presence of some recently formed stars. After comparing the characteristic timescales of star formation with the age of IC443, we conclude that the protostellar objects and YSOs candidates are not triggered by IC443. For the age of stellar winds shell, we have performed calculation on the basis of stellar winds shell expansion model. The results and analysis suggest that the formation of these stars may be triggered by the stellar winds of IC443 progenitor.

*Subject headings:* ISM: clouds — ISM: individual (IC443) — ISM: molecules — stars: formation — supernova remnants

## 1. INTRODUCTION

The progenitor of core-collapse supernova remnant (SNR) is high-mass star (with masses $> 8\ M_\odot$). High-mass star could dramatically impact the surrounding environment through stellar winds at their earliest evolutionary stage and supernova explosion at their last evolutionary stage. When SNR shocks expand into the surrounding molecular clouds (MCs), the local density of the MCs increases then some condensed MCs will collapse. The timescale of collapsed phase is $\sim 10^5$ yr (Lefloch & Lazareff 1994), while protostellar objects occur in a period on the order of $10^6$ yr (Reynoso & Mangum 2001). In principle, galactic density waves, the shocks of cloud-cloud collision (e.g., Wang et al. 2004; Xin et al. 2008), energetic stellar winds (Jankes et al. 1992) and SNR shocks (Frail et al. 1994; Reynoso & Mangum 2000; Melioli et al. 2006) all could induce collapse of MCs to protostellar objects. Therefore, the interaction between SNR and MCs is expected to play an essential part in the process of star formation.

SNR IC443, at a distance of 1.5 kpc (Fesen 1984), is located in the Gem OB1 association (Heiles 1984). A pulsar is associated with IC443 (Olbert et al. 2001; Gaensler et al. 2006), suggesting a core-collapse origin. In addition, the age of IC443 is 3000 $\sim$ 30000 yr (Petre et al. 1988; Olbert et al. 2001; Lee et al. 2008). The prior observations suggested that IC443 may be interacting with the surrounding MCs (DeNoyer 1979; Musfson et al. 1986; Dickman et al. 1992; van Dishoeck & Jansen 1993). Braun & Strom (1986) suggested that IC443 has evolved in a cavity of wind-blown bubble, possibly formed by the pre-SN evolution, and recently started to interact with the cavity shell. The radius of the molecular shell is 7.1 pc (Lee et al. 2008). Three 1720-MHz OH masers were detected around the molecular shell of IC443 (Hewitt et al. 2006). The 1720-MHz OH masers have been identified as one sure signature of SNR-MCs interaction (Frail et al. 1994; Yusef-Zadeh et al. 2003). However, whether new generation of star formation triggered by IC443 is still poorly understood.

In this paper, we have performed CO $J = 2-1$ and $J = 3-2$ observations toward IC443. The observations cover for the first time the whole area of IC443 in these frequencies. To investigate the impact of SNR IC443 on the process of star formation, we used *IRAS* Point Source Catalog and 2MASS near-infrared database to select protostellar objects and young stellar objects (YSOs) candidates (including CTTSs and HAeBe stars). The observations are described in §2, and the results are presented in §3. In §4, we discuss how our data lend supportive evidence of triggered star formation in the interacting region. The conclusions are summarized in §5.

## 2. Observations

The mapping observations of IC443 were made in CO $J = 2-1$ and CO $J = 3-2$ lines using the KOSMA 3m telescope at Gornergrat, Switzerland in March 2002. The half-power beam widths of the telescope at observing frequencies of 230.538 GHz and 345.789 GHz, are $130''$ and $80''$, respectively. The pointing and tracking accuracy is better than $10''$. The DSB receiver noise temperature was about 120 K. The medium and variable resolution acousto optical spectrometers have 1501 and 1601 channels, with total bandwidth of 248 MHz and 544 MHz, and equivalent velocity resolution of 0.21 km s$^{-1}$ and 0.29 km s$^{-1}$, respectively. The beam efficiency $B_{\rm eff}$ is 0.68 and 0.72 at 230 GHz and 345 GHz, respectively. The forward efficiency $F_{\rm eff}$ is 0.93. The mapping was done using on-the-fly mode with a $1' \times 1'$ grid. The data was reduced using the CLASS (Continuum and Line Analysis Single-Disk Software) and GREG (Grenoble Graphic) software.

The 1.4 GHz radio continuum emission data were obtained from the NRAO VLA Sky Survey (NVSS; Condon et al. 1998).

## 3. RESULTS

### 3.1. Molecular emission

Figure 1a presents the integrated intensity maps of CO $J = 2-1$ and CO $J = 3-2$, overlapping the 1.4 GHz radio continuum emission. In Figure 1a, CO $J = 2-1$ and CO $J = 3-2$ maps appear as incomplete shell, which may be produced by energetic events near the center, such as strong stellar winds and/or supernova explosions. Cloud B-G are clearly identifiable as local integrated intensity peaks around IC443. IC443 also shows an incomplete radio shell morphology on the whole. Each cloud here has been designated alphabetically, following the naming of DeNoyer (1979) and Huang et al. (1986). Cloud B, C, D and E are associated with the 1.4 GHz radio continuum emission of IC443. CO $J = 2-1$ and CO $J = 3-2$ molecular emission is significantly stronger in cloud B, C and G. However, the molecular emission from cloud A and H are too weak to be detected at the signal-to-noise ratio of our survey. Three 1720-MHz OH masers, detected in the radio shell of IC443 (Hewitt et al. 2006), are located close to the peak of cloud C, D and G, respectively. Figure 1b shows CO $J = 3-2$ spectra at the peak position of cloud B-G, respectively. Line profile of cloud B and C is greatly broadened in the blue wing, while line profile of cloud F is greatly broadened in the red wing. For cloud C, CO emission in velocity intervals -60−2 km s$^{-1}$ is well associated with a broad component of molecular hydrogen gas at -30 km $s^{-1}$ (Rosado et al. 2007). The fitted and derived parameters for cloud B-G are summarized in Table 1.

The different transitions of CO trace different molecular environment. In order to obtain the integrated intensity ratio of CO $J = 3-2$ to CO $J = 2-1$ (R$_{I_{\rm CO(3-2)}/I_{\rm CO(2-1)}}$), we convolved the 80″ resolution of CO $J = 3-2$ data with an effective beam size of $\sqrt{130^2 - 80^2} = 102''$. The integrated intensities were calculated for CO $J = 3-2$ line in the same velocity range as for CO $J = 2-1$ line. The integrated range is from -60 to 28 km s$^{-1}$. Figure 2 shows the distribution of R$_{I_{CO(3-2)}/I_{CO(2-1)}}$ (color scale) overlaid with the distribution of CO $J = 3 - 2$ line integrated intensity (contours). The maximum ratio of cloud C is 3.40, which is located at R.A.=06$^h$17$^m$42$^s$ (J2000), Dec=+22°22′34″ (J2000). For cloud G, the maximum ratio (3.18) is located at R.A.=06$^h$16$^m$35$^s$ (J2000), Dec=+22°32′34″ (J2000). The ratio value for the whole molecular cloud is between 0.79 and 3.40. The average value is 1.58, which is much higher than typical value (0.55) for molecular clouds in the Galactic disk (Sanders et al. 1993) and value (0.69) for the normal MCs in M33 (0.69 Wilson et al.1997), and even higher than value (0.8) in the starburst galaxies M82 (Guesten et al. 1993).

We assume local thermodynamical equilibrium (LTE) for the gas and optically thick condition for CO $J = 3 - 2$ line to use the relation $N_{H_2} \approx 10^4 N_{CO}$ (Dickman 1978). The optical depth of $\tau = 3.1$ is given by braun & strom (1986). The column density is estimated as (Garden et al. 1991)

$$N_{CO} = 1.89 \times 10^{15}(T_{ex} + 0.92)\exp(16.6/T_{ex}) \int T_{mb} dv \ \text{cm}^{-2}, \tag{1}$$

The excitation temperature of CO $J = 3 - 2$ line, $T_{ex}$, is estimated following the equation $T_{ex} = 16.6/\ln[1 + 1/(T_{mb}/16.6 + 0.0024)]$, where $T_{mb}$ is the corrected main beam temperature. If cloud cores are approximately spherical in shape, the mean H$_2$ number density is $N(H_2) = 1.62 \times 10^{-19} N_{H_2}/L$, where $L$ is the cloud core diameter in parsecs (pc). The mean diameter is 1.75 pc for all the MCs. Furthermore, their mass is given by $M_{H_2} = N_{H_2}\frac{1}{6}\pi L^3 \mu_g m(H_2)$ (Garden et al. 1991), where $\mu_g$=1.36 is the mean atomic weight of the gas, and $m(H_2)$ is the mass of a hydrogen molecule.

### 3.2. Infrared emission

*IRAS* (IR) point source is good signpost of recent star formation. In order to explore a causal relationship between IC443 and star formation, we have searched for protostellar candidates in the *IRAS* Point Source Catalog that fulfill the following selection criteria (Jankes et al. 1992): (1) $F_{100} \geq 20$ Jy, (2) $1.2 \leq F_{100}/F_{60} \leq 6.0$, (3) $F_{25} \leq F_{60}$, and (4) $R_{IRAS} \leq 2 \cdot R_{\mathrm{radio}}$, where $F_{25}, F_{60}$ and $F_{100}$ are the infrared fluxes at three IR bands ($25\mu$m, $60\mu$m and $100\mu$m), respectively. The first criterion selects only strong sources. The second and third discriminate against cold IR Point sources probably associated with cool stars, planetary nebulae, and cirrus clumps. While the fourth guarantees that the search diameter include the complete surface of IC443. However, these selection criteria may include not only protostars but also dust heated by shocks. Dozen IR point sources were found in a search circle around IC443 within 25′ radius centered at R.A.=$06^{\mathrm{h}}17^{\mathrm{m}}13^{\mathrm{s}}.0$ (J2000), Dec=+22°37′10″ (J2000). The coordinates are adopted as the explosion center (Lee et al. 2008). The name and coordinates of these IR point sources are listed in Table 2. Figure 3 shows that the IR sources are associated quite well with the CO molecular shell. IR 4 and IR 10 are located on the peak position of CO molecular shell. Infrared luminosity (Casoli et al. 1986) and dust temperature (Henning et al. 1990) are expressed respectively as,

$$L_{\mathrm{IR}} = (20.653 \times F_{12} + 7.538 \times F_{25} + 4.578 \times F_{60} + 1.762 \times F_{100}) \times D^2 \times 0.30, \quad (2)$$

$$T_{\mathrm{d}} = \frac{96}{(3+\beta)\ln(100/60) - \ln(F_{60}/F_{100})}. \quad (3)$$

Where $D$ is the distance from the sun in kpc. The emissivity index of dust particle ($\beta$) is assumed to be 2. The calculated results are presented in Table 2.

To further look for primary tracers of star formation activity in the vicinity of IC443, we used the 2MASS All-Sky Point source database in the near-infrared $J(1.25~\mu$m), $H(1.65$ $\mu$m) and $K_s$(2.17 $\mu$m) bands, with a signal-to-noise ratio (S/N) greater than 10. The 2MASS database provides a good opportunity to select young stellar objects (YSOs) candidates associated with or even embedded in molecular clouds. From the database, we selected 10272 near-infrared sources in a search circle around IC443 within 25' radius. Figure 4 shows the $(H-K_s)$ versus $(J-H)$ color-color (CC) diagram. The two solid curves represent the location of the main sequence and the giant stars, respectively (Bessell & Brett 1988). The parallel black dashed lines are reddening vectors. The blue dashed line is $(J-H) - 1.7(H-K_s) + 0.450 = 0$, the blue solid line is $(J-H) - 0.493(H-K_s) - 0.439 = 0$, the red dashed line is $(J-H) - 1.7(H-K_s) + 1.400 = 0$ and red solid line is $(J-H) = 0.2$. 2MASS sources are classified into three regions: cool giants, normally reddened stars and infrared excess sources. Because YSOs have near-infrared excess (Lada & Adams 1992), YSOs candidates are located at the rightmost of reddened vectors (Weintraub et al. 1996; Hanson et al. 1997; Paron et al. 2009). Based on this criteria we find 1666 YSOs candidates in the mentioned region. Figure 5a presents the spatial distribution of these selected YSOs candidates. It is clear that the YSOs candidates are not symmetrically distributed in the whole selected regions, and are mostly concentrated in cloud C and G. Regarding the geometric distribution of the YSOs candidates, we plot the map of the star surface density, which was obtained by counting all the YSOs candidates with a detection in the $J, H$ and $K_s$ bands in squares of $4' \times 4'$, as shown in Figure 5b. From Figure 5b we can see that there are clear signs of clustering toward the interacting regions. The number of the YSOs candidates increases when going from the border of clouds to the center. The existence of YSOs candidates may indicate star formation activity.

Because the selected YSOs candidates exhibit different near-IR excess, we then separated these YSOs candidates into three classes based on criteria developed by Lee & Chen (2007), including low-mass Classical T Tauri stars (CTTSs) and intermediate-mass Herbig Ae/Be (HAeBe) stars and other YSOs candidates. Generally, CTTSs are in an earlier evolutionary stage, while HAeBe stars exhibit a stronger near-IR excess than CTTSs. In Figure 4, CTTS candidates lie between the rightmost black dashed line and the blue dashed line, and above blue solid line; HAeBe star candidates lie between the blue dashed line and the red dashed line, and above red solid line. In this way, we found a total of 154 CTTS candidates and 419 HAeBe star candidates. Figure 6 shows that the spatial distribution of selected CTTS and HAeBe star candidates. From Figure 6 we clearly see that the CTTS and HAeBe star candidates are concentrated and grouped around the CO molecular shell, respectively. Because these CTTS and HAeBe star candidates are grouped together, they are likely to be CTTSs and HAeBe stars.

## 4. DISCUSSIONS

### 4.1. IC443 and MCs

The morphology of IC443 in the 1.4 GHz radio continuum emission is coincident very well with the surrounding MCs except for the northern region where there are no MCs detected. We suggest that the shocks interacting with the MCs in northern region are strong enough so as to blow away the MCs. On the other hand, it may be that the CO emission of MCs in northern region is too weak to be detected at the signal-to-noise ratio of our survey. CO molecular emission from the MCs shows a shell morphology on the whole. The broadened line wings were detected in molecular shell. The radio continuum emission also presents a a shell morphology, which may arise from the SNR shocks that is blocked by surrounding condensed MCs (Frail & Mitchell 1998). Three 1720-MHz OH masers were detected around the radio shell of IC443 (Hewitt et al. 2006). Theoretical studies suggest that 1720-MHz OH masers are more efficiently pumped in hot, condensed shocks clouds. The masers may be excited when a nondissociative shocks, produced by the interaction of IC443 with the surrounding MCs, propagates into the sufficiently condensed

MCs. Together, these observations strongly indicate that the MCs is interacting with IC443. The maximum value of $R_{I_{\rm CO(3-2)}/I_{\rm CO(2-1)}}$ in cloud C and G are higher than that in cloud B, D, E and F. We consider that the positions of cloud C and G are the strongly interacting regions. The comparison with previously published observations reveals that $R_{I_{\rm CO(3-2)}/I_{\rm CO(2-1)}}$ for the MCs associated with IC443 are systematically larger than that for the MCs in the Galactic disk (Sanders et al. 1993), the M33 (Wilson et al. 1997), and the starburst galaxies M82 (Guesten et al. 1993). High $R_{I_{\rm CO(3-2)}/I_{\rm CO(2-1)}}$ in starburst galaxies may be due to unusual conditions in these dense, hot regions (Aalto et al. 1997), while for normal MCs the most likely explanation is a significant contribution to the CO emission by low column density material (Wilson & Walker 1994). When supernova (SN) expands into the surrounding parental MCs, the shocks resulting from the interaction of SNR with the MCs can heat gases. As the temperature of gases increases, the line opacities decrease as the upper $J$ levels become more populated. Depopulating the $J = 2$ and $J = 3$ levels lead to the $R_{I_{\rm CO(3-2)}/I_{\rm CO(2-1)}}$ increase. High $R_{I_{\rm CO(3-2)}/I_{\rm CO(2-1)}}$ ($\sim$1.58) in the interacting region may imply that shocks have driven into MCs, and have impacted on the temperature and chemistry of MCs. We consider that high $R_{I_{\rm CO(3-2)}/I_{\rm CO(2-1)}}$ is also identified as one good signature of SNR-MCs interaction system. From Table 1, the total mass of MCs is $9.26 \times 10^3$ $M_\odot$; The value indicates that the whole molecular cloud associated with IC443 may be a giant molecular cloud (GMC) ($\geq 10^4 M_\odot$ for GMC).

### 4.2. Recent star formation in the MCs

The superposition of $IRAS$ (IR) 4 and IR 10 with cloud peaks and the association of other IR sources with CO molecular shell are very unlikely to be by chance. If IR sources simply mark intensity peaks in CO molecular shell, it will be difficult to explain why we did not see IR sources at the other unshocked regions. We suggest that the selected IR sources represent the signatures of protostars. The protostars formation may be triggered by shocks. From Table 2, all the IR sources are with $L_{IR} < 10^3 L_\odot$, thus all the selected IR sources are considered to be low-mass protostars. We suggest that the shocks interacting with the MCs may be strong enough to tear up the condensed GMC.

From the distribution of YSOs candidates in the interacting regions, we identified two young star clusters, cluster C and G. Since these redder NIR sources are clustered, it is unlikely that they all are just reddened foreground and background stars. It is more likely that most of YSOs candidates are YSOs, physically associated with the interacting regions between IC443 and MCs. Because the distribution of YSOs candidates is grouped and elongate around CO molecular shell, it strongly support triggering star formation. In addition, low-mass CTTSs candidates and intermediate-mass HAeBe stars candidates are grouped together in the strongly interacting regions, indicating that these stars more likely to be formed by triggering.

Lefloch & Lazareff (1994) mentioned that star formation may undergo a collapse phase of timescale of $\sim 10^5$ yr; Condensed molecular matter can form a protostar after the matter has cooled, which occurs in a period on the order of $10^6$ yr; While the age of CTTS is $10^5 \sim 10^7$ yr. Such periods are far longer than the age of IC443, which is $3000 \sim 30000$ yr (Petre et al. 1988; Olbert et al. 2001; Lee et al. 2008). Hence the formation of the protostars and YSOs candidates (including CTTSs and HAeBe Stars) may not be triggered by the shocks of SNR IC443. Braun & Strom (1986) suggested that supernova explosion of IC443 had occurred in pre-existing cavity blown by the stellar wind of high-mass progenitor. The spectral type of IC443 progenitor is O or B (Heiles 1984). Applying the equation $R_W = 4.3 \times 10^{-10} (L_W/n_0)^{1/5} t_W^{3/5}$ (Castor et al. 1975), where $n_0$ [cm$^{-3}$] is medium density, $L_W$ [ergs/s] is wind luminosity, and $R_W$ [pc] is MC shell radius. The stellar winds parameters for O8I or B1III are taken from model computation (Junkes et al. 1992). The maximal bubble radius created by an O8I (B1III) star is $R_W = 12$ pc (22 pc) for the mean density of $n_0 = 5.29 \times 10^3 \text{cm}^{-3}$, however, the IC443 bubble radius is 7.1 pc (Lee et al. 2008). Therefore, IC443 progenitor can provide enough energy for the formation of a CO molecular shell with the observed values for radius. Furthermore, we derive the time intervals for the stellar winds of a young star to blow up a bubble with $R_W = 7.1$ pc in a medium with density $n_0 = 5.29 \times 10^3 \text{cm}^{-3}$. The results are $t_W = 1.2 \times 10^6$ yr (O8) or $t_W = 1.0 \times 10^6$ yr (B1). This can be compared with the main-sequence (MS) lifetime of such stars ($2.0 \times 10^6$ yr for O5 stars). In the course of high-mass stars evolution the stellar winds go through three phases (Dwarkadas 2007). The first and longest lasting is the main-sequence stage, which lasts for about $4.6 \times 10^6$ yr. At the end of the MS phase the star swells up immensely in size to become a red supergiant (RSG). Since the RSG phase lasts about $2.5 \times 10^5$ yr. At the end of the RSG phase, the star sheds its H envelope and becomes a Wolf-Rayet (W-R) star. Such periods of stellar winds are sufficient for triggering the collapse of molecular cloud, and to form protostars and YSOs candidates. We further conclude that the formation of the protostars and YSOs candidates may be triggered by the stellar winds from the high-mass progenitor of IC443. The distribution of YSOs candidates may show a shell morphology (see Figure 5b ), which is needed to be confirmed further. Moreover, the newly forming stars triggered by the shocks of IC443 may be in the phase of deeply embedded protostars. In future, we need to perform much higher angle resolution submillimeter observations to detect them.

## 5. SUMMARY

We have presented large-area map of molecular clouds (MCs) in the vicinity of SNR IC443 in CO $J = 2-1$ and CO $J = 3-2$ lines. The shell-sharped morphology, the broadened emission lines, and the associated three OH masers further suggest that these MCs are interacting with IC443. The integrated CO line intensity ratio ($R_{I_{CO(3-2)}/I_{CO(2-1)}}$)($\sim 1.58$) for the whole molecular cloud exceeds previous measurements of individual Galactic molecular cloud, implying that shocks have driven into MCs. We suggest that high $R_{I_{CO(3-2)}/I_{CO(2-1)}}$ is also identified as one good signature of SNR-MCs interacting system. The whole molecular cloud associated with IC443 is most probably giant molecular cloud, with the total mass of $9.26 \times 10^3$ M$_\odot$. The selected protostellar objects and young stellar objects (YSOs) candidates (including CTTSs and HAeBe stars) are concentrated and grouped around the interacting regions. It provides strong signpost for ongoing star formation. Comparing the age of IC443 and the timescales for the stellar winds of IC443 progenitor with the characteristic timescales for star formation, we conclude that the protostellar objects and YSOs candidates may not be triggered by SNR IC443, but triggered by the stellar winds of IC443 progenitor. Moreover, the newly forming stars triggered by the shocks of IC443 may be in the phase of deeply embedded protostellar.

We are very grateful to the anonymous referee for his/her helpful comments and suggestions. This work was supported by the National Natural Science Foundation of China under Grant No.10473014.

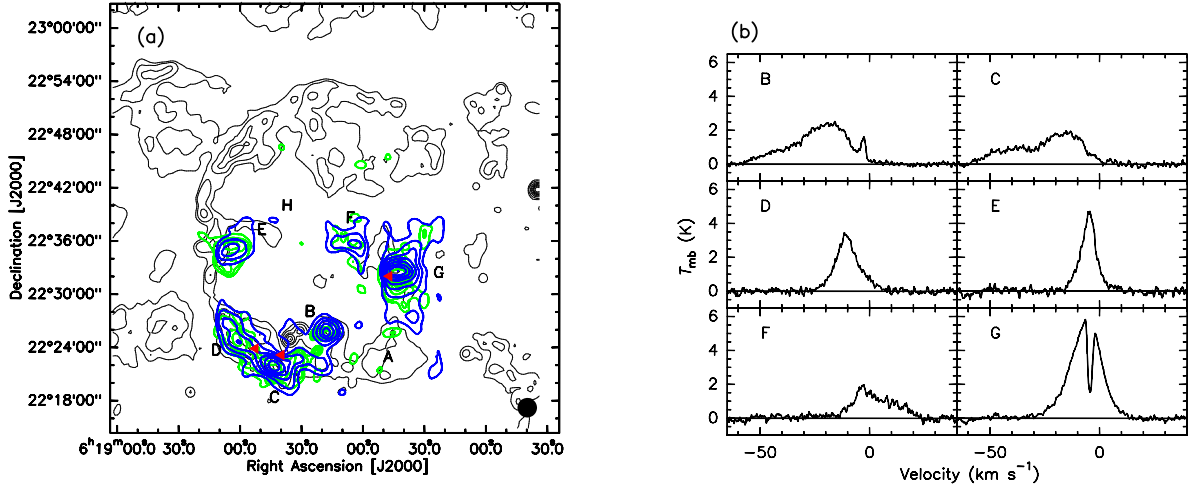

Fig. 1.— (a) CO $J = 2 - 1$ (green contours) and $J = 3 - 2$ (blue contours) intensity maps integrated from -60 to 28 km s$^{-1}$, overlaid on the NVSS 1.4 GHz radio continuum emission contours (contour levels are 0.002, 0.008, 0.032, 0.064, 0.128, 0.181, 0.256, 0.312 and 0.512 Jy beam$^{-1}$). The green and blue contours start at a 30% level of the peak in steps of every 10%. OH masers are shown with red triangle (Hewitt et al. 2006). The beam is shown with the black solid circle in the lower right corner. (b) CO $J = 3 - 2$ spectra of at the peak position of B-G cloud clumps. Letter A, B, C, D, E, F, G and H indicate the different MCs (Denoyer 1979 and Huang et al. 1986).

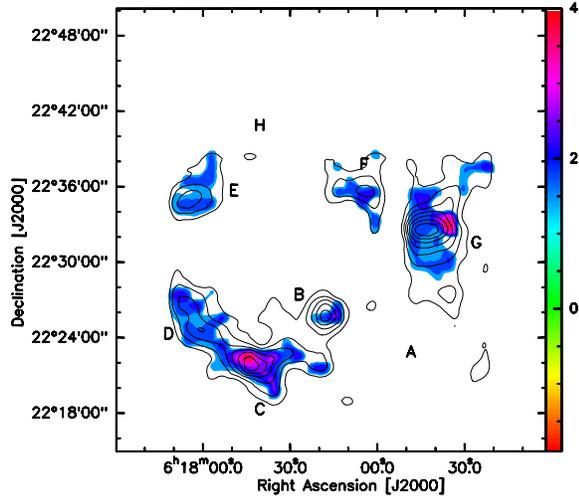

Fig. 2.— CO $J = 3 - 2$ intensity map are superimposed on the line intensity ratio map (color scale), the line intensity ratios ($R_{I_{CO(3-2)}/I_{CO(2-1)}}$) range from 0.79 to 3.40 by 0.34. The wedge indicates the line intensity ratios scale. Letter A, B, C, D, E, F, G and H indicate the different MCs (Denoyer 1979 and Huang et al. 1986 ).

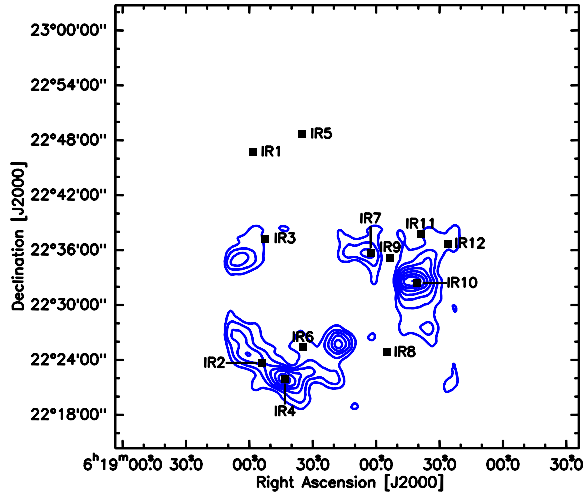

Fig. 3.— Positions of IR Point sources relative to the CO $J = 3 - 2$ shell around IC443. The IR sources are labeled as the filled square symbols.

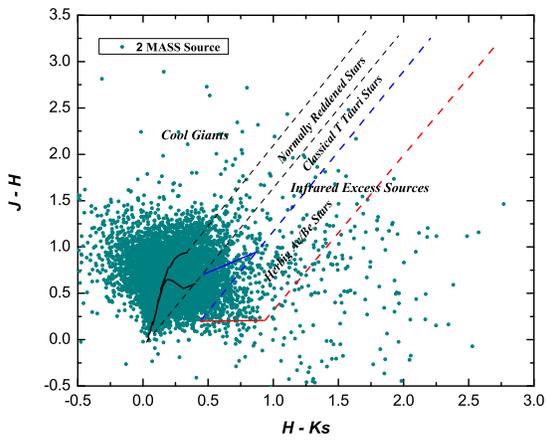

Fig. 4.— Color-color diagram of NIR sources in the vicinity of IC443. The two solid curves represent the location of the main sequence and the giant stars, respectively. The parallel dashed lines (black) are reddening vectors. The plot is classified into three regions: cool giants, normally reddened stars and infrared excess source. CTTS candidates lie between the rightmost black dashed lines and the blue dashed lines, and above blue solid line. HAeBe star candidates lie between the blue dashed lines and the red dashed lines, and above red solid line.

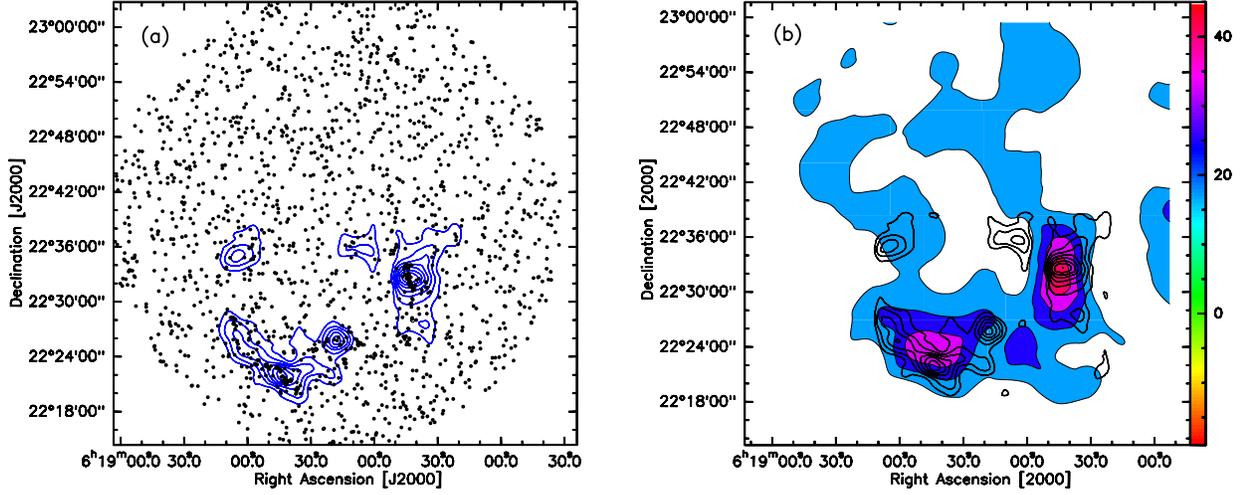

Fig. 5.— (a) Positions of YSOs relative to the CO $J = 3 - 2$ MCs ( blue contours) around IC443. YSOs are labeled as the solid circles. (b) CO $J = 3 - 2$ intensity map are superimposed on the stellar-surface density map of all YSOs detected in the J, H and K. Contours range from 15 to 45 stars $(4\text{arcmin})^{-2}$ in steps of 10 stars $(4\text{arcmin})^{-2}$. $1\sigma$ is 5 $(4\text{arcmin})^{-2}$ (background stars). The wedge indicates the line intensity ratios scale, its units are stars $(4\text{arcmin})^{-2}$.

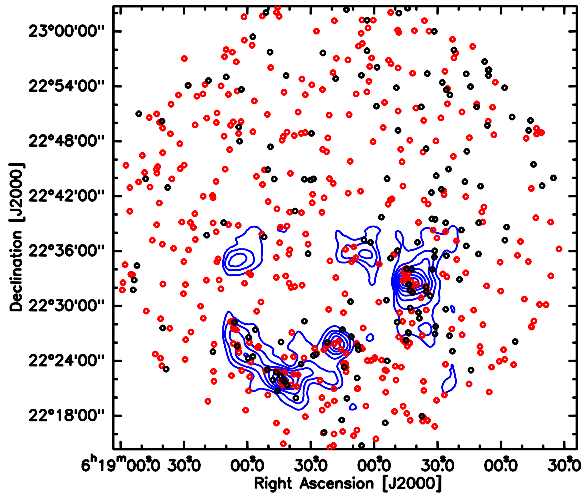

Fig. 6.— Positions of CTTS and HAeBe Star candidates relative to the CO $J = 3 - 2$ MCs ( blue contours) around the IC443. The CTTSs are labeled as the black circles, the HAeBe Stars are labeled as the red circles.

Table 1: Observed and derived parameters for the CO clouds.

| Parameter | Cloud B | Cloud C | Cloud D | Cloud E | Cloud F | Cloud G |
|---|---|---|---|---|---|---|
| $T_{\text{peak}}(\text{K})$ | 2.54 | 2.39 | 3.40 | 4.83 | 1.88 | 5.92 |
| $T_{\text{ex}}(\text{K})$ | 8.27 | 8.06 | 9.42 | 11.19 | 7.32 | 12.47 |
| $N_{\text{CO}}(10^{18}\ \text{cm}^{-2})$ | 7.58 | 8.41 | 4.33 | 3.45 | 3.77 | 7.28 |
| $N_{\text{H}_2}(10^{22}\ \text{cm}^{-2})$ | 7.58 | 8.41 | 4.33 | 3.45 | 3.77 | 7.28 |
| $N(\text{H}_2)(10^{3}\ \text{cm}^{-3})$ | 7.02 | 7.79 | 4.01 | 3.19 | 3.49 | 6.24 |
| $M_{\text{H}_2}(10^{3}\ \text{M}_\odot)$ | 1.98 | 2.20 | 1.13 | 0.90 | 0.99 | 2.06 |

Table 2: Selected IR point sources near IC443: IR flux densities, dust temperatures and IR luminosities.

| Name | Source | RA (h m s) | DEC (° ′ ″) | $F_{12}$ [Jy] | $F_{25}$ [Jy] | $F_{60}$ [Jy] | $F_{100}$ [Jy] | $T_d$ [K] | $L_{IR}$ [$L_\odot$] |
|---|---|---|---|---|---|---|---|---|---|
| IR1 | IRAS 06149+2247 | 06 17 58.84 | 22 46 46.05 | 0.37 | 0.49 | 12.71 | 24.26 | 30.00 | 80.00 |
| IR2 | IRAS 06148+2224 | 06 17 54.71 | 22 46 46.05 | 0.41 | 0.36 | 10.3 | 47.03 | 23.57 | 99.45 |
| IR3 | IRAS 06148+2238 | 06 17 53.32 | 22 37 15.44 | 1.13 | 0.56 | 8.52 | 31.06 | 24.95 | 86.47 |
| IR4 | IRAS 06147+2223 | 06 17 43.67 | 22 21 51.14 | 0.65 | 0.53 | 13.62 | 65.23 | 23.30 | 137.32 |
| IR5 | IRAS 06145+2249 | 06 17 35.28 | 22 48 44.77 | 0.25 | 0.23 | 8.84 | 24.28 | 26.93 | 63.77 |
| IR6 | IRAS 06145+2226 | 06 17 35.05 | 22 25 19.77 | 0.42 | 0.37 | 7.45 | 41.26 | 22.50 | 83.50 |
| IR7 | IRAS 06140+2236 | 06 17 02.59 | 22 35 39.13 | 0.25 | 0.39 | 12.83 | 38.30 | 26.32 | 94.28 |
| IR8 | IRAS 06138+2225 | 06 16 54.84 | 22 24 48.69 | 0.29 | 0.29 | 9.17 | 41.26 | 23.65 | 86.46 |
| IR9 | IRAS 06138+2236 | 06 16 53.18 | 22 35 07.82 | 0.25 | 1.23 | 10.62 | 42.51 | 24.36 | 97.08 |
| IR10 | IRAS 06136+2233 | 06 16 40.22 | 22 32 22.75 | 0.44 | 0.40 | 6.72 | 30.24 | 23.66 | 68.03 |
| IR11 | IRAS 06136+2238 | 06 16 38.44 | 22 37 45.89 | 0.28 | 0.47 | 10.62 | 39.02 | 24.90 | 89.15 |
| IR12 | IRAS 06134+2237 | 06 16 25.62 | 22 36 42.82 | 0.53 | 0.48 | 5.46 | 24.71 | 23.63 | 59.15 |